\documentstyle[prl,aps,multicol,epsf]{revtex}
\renewcommand{\narrowtext}{\begin{multicols}{2}
\global\columnwidth20.5pc} 
\renewcommand{\widetext}{\end{multicols}
\global\columnwidth42.5pc} \multicolsep = 8pt plus 4pt minus 3pt

\newcommand{\be}{\begin{equation}}
\newcommand{\ee}{\end{equation}}
\begin{document}

\title{Spectral Functions of 1D Hubbard Rings with Varying 
Boundary Conditions}

\author{R.N. Bannister, N. d'Ambrumenil}
\address{Department of Physics, University of Warwick, Coventry, CV4 7AL}

\maketitle


\begin{abstract}
We study the effect of varying the boundary condition on
the spectral function of a finite 1D Hubbard chain, which
we compute using direct (Lanczos) diagonalization of the Hamiltonian.
By direct comparison with the two-body response
functions and with the exact solution
of the Bethe Ansatz equations, we can identify both
spinon and holon features in the spectra.
At half-filling the spectra have the well-known 
structure of a low energy holon band and its shadow---which span 
the whole Brillouin zone---and a spinon band present for momenta 
less than the Fermi momentum.  New features related to the twisted 
boundary condition are cusps in the spinon band.  We show that the 
spectral building principle, adapted to account for both the finite 
system size and the twisted boundary condition, describes the 
spectra well in terms of single spinon and holon 
excitations. 
We argue that these finite-size effects
are a signature of spin-charge separation
and that their study should help establish the
existence and nature of spin-charge separation
in finite-size systems.

\end{abstract}

\bigskip
\bigskip

\pacs{PACS Numbers: 74.20.-z, 74.20Mn, 75.25.Dw} 

\narrowtext

\section{INTRODUCTION}

The nature of both the ground and excited states for
the Hubbard model (HM) has been the theme of many 
recent studies.  The HM itself is important not 
only because of its simplicity but because it is 
believed to include many of the 
principle features of strong electron correlations
in a wide range of real materials. These include
the 2D cuprate 
high-$T_c$ materials \cite{Anderson} and in 1D, compounds such as ${\rm 
TTF}$-${\rm TCNQ}$ \cite{Maldague} and ${\rm SrCuO}_2$ \cite{Kim1}. 

The eigenvalues can be found exactly for the 1D HM from the 
Bethe Ansatz (BA) equations \cite{Lieb}.  
The corresponding eigenstates can be interpreted in terms of
separate populations of spin and charge excitations. 
The separate dynamics of spin and charge 
(spin-charge separation) together with 
other anomalous properties 
of the HM in 1D \cite{Schultz,Brech} 
are now considered to be the standard signatures of 
non-Fermi (or Luttinger) Liquid behavior in 1D systems
and is found in other exactly soluble 1D
models like the
Tomonoga-Luttinger model \cite{Haldane81,Carmelo91}.

The exact BA solution is not available in 
higher-dimensional
versions of the model nor even in 1D when `extended' effects such as 
next nearest neighbor hopping are included. Instead,
one has to resort to
finite-size system calculations and approximate
methods. One of the problems with finite-size system
calculations is that the restriction  to small
systems can make information about the nature of low-energy
excitations in the thermodynamic limit hard
to extract. Even so exact diagonalizations are 
an increasingly important method for studying 
strongly correlated electron systems
\cite{Riera99}.

Here we investigate how finite-size
effects directly reflect the nature of the 
underlying excitations. In particular we look at how
spin-charge separation manifests itself in the variation
of the spectral function as a function of
the boundary condition. We work
with the 1D HM where the characterization
of results of direct 
numerical diagonalization (in terms of spinons and holons) 
can be independently verified
by comparing with the exact BA solution.

There has already been considerable interest
in how the HM behaves as the boundary conditions are 
changed by introducing a phase twist, $\delta$, 
in the periodic boundary conditions 
(PBC), {\it ie} $c^\dagger_{i+N}=e^{-i\delta}c^\dagger_i$.
This corresponds to threading the loop formed by the $N$
sites in the chain with the magnetic flux
$\Phi_0\delta/2\pi$, where $\Phi_0$ is the flux quantum
and the $c^\dagger_i$ create electrons in orbitals at 
site $i$. One consequence of the phase twist is 
to shift all allowed momenta by an amount $\delta/N$ yielding 
the allowed momenta:
\begin{equation}\label{kset}
k=\frac{2n\pi+\delta}{N},
\end{equation}
for $n$ integer.  
When an electron is added or removed, the many body 
system changes its total momentum by an amount corresponding
to one of the allowed values in (\ref{kset}). 
By varying $\delta$ we can, in effect, simulate a continuously
varying momentum in a finite system.  

The ground state energy of the 1D HM has an interesting structure 
when plotted 
as a function of $\delta$ \cite{Yu,Kusmartsev,Schofield91}.  
For large $U$, 
there are oscillations in the ground state energy which are often 
found to vary 
with more than one period.  These oscillations---which may include 
periodicities 
of $1$, $1/N_e$ and $N_\downarrow/N_e$ (for a total of $N_e$ electrons 
present, 
$N_\downarrow$ of them being spin-down)---give rise to cusps in the energy 
variation for a finite system.  These effects are accounted for in 
terms of 
transitions mainly in the sea of spin particles as $\delta$ is changed 
\cite{Yu,Kusmartsev}.  The ground state energy is found to be consistent 
with a hole in the spin sea which traverses from one 
Fermi point to the other as the boundary condition
is varied. (Since the spin `excitations' have zero energy,
the spin sector can be used by the system as a momentum
bank charging no `interest'.)

To gain a quantitative picture of excitations in the 1D HM, various
authors have used different approximations 
and direct diagonalizations.  By using Lanczos diagonalization, 
Kim and collaborators \cite{Kim1,Kim2} calculated the photoemission 
spectral 
function, $A(k,\omega)$, for a finite half-filled system with fixed 
PBC (they 
actually considered the {\it t-J} model).  Their result, which is 
essentially 
the same as that of the strongly correlated HM, consists of spectral 
weight belonging to two bands, which they identified with separated
spin and charge degrees of freedom.  
The `charge band' (of width $4|t|$ and present in 
the whole Brillouin zone) traces out the dispersion of the 
so-called holon excitations. 
The `spin band' (of width $\pi J/2$ for $J=4t^2/U$) exists only for 
$|k|<k_F$ $(k_F=\pi /2)$ and is associated with an excitation of the 
corresponding spinon population.  
This structure of the spectral function is 
a natural consequence of spin-charge separation and can be explained 
through a 
semi-intuitive picture which has been termed the `spectral building 
principle' 
(SBP) \cite{Eder1,Kim1} (see section IV below).

In this paper we 
study the spectral functions of small 1D systems ($N=12$) 
at half-filling and in 
the presence of variable twisted PBC's. We show how the spectral building
principle can be adapted to account for the features that appear in
the spinon bands and suggest that these features provide
a natural test in finite system calculations for spin-charge separation.
 The structure of this paper is as 
follows:  In section II we outline the procedure which we adopt to 
calculate and 
analyze the spectral functions of the 1D HM with twisted PBC.  Our 
results are 
shown in section III.  In section IV we review and adapt the SBP to help 
interpret our spectra.  Finally in section V we summarize our findings 
and 
speculate how the SBP could have a bearing on systems other than the 
half-filled 
HM in 1D.

\medskip
\section*{II. Procedure}
We compute the spectral functions using the method used by Tohyama and 
Maekawa \cite{Tohyama}.  For a given $\delta$ we compute the spectral 
function 
$A(k,\omega)$,
\begin{equation}\label{photoemission}
A(k,\omega)=\frac{1}{N}\sum_m|\langle\psi^{N_e-1}_m|c_{k\downarrow}
|\psi^{N_e}_0
\rangle|^2\delta(\omega+[E^{N_e-1}_m-E^{N_e}_0])
\end{equation}
at those momenta consistent with the choice of $\delta$.  
We use the standard 
Lanczos recursion method \cite{Dagotto}.  In eq. (\ref{photoemission}), 
$|\psi^{N_e}_m\rangle$ is the $m$th energy eigenvector 
(with energy $E^{N_e}_m$) 
of the $N_e$-electron system (at half-filling 
there are equal populations of 
each spin species).  All wavefunctions and energies are 
dependent upon $\delta$. 
 Following \cite{Tohyama} we analyze the peaks in the spectral 
function by computing the complementary 
(fixed particle number) dynamic spin and charge correlation 
functions defined respectively as,
\begin{eqnarray}
S(q,\omega)&=&\frac{1}{N}\sum_m|\langle\psi^{N_e-1}_m|S(q)|
\psi^{N_e-1}_0\rangle |^2 \nonumber \\
 &  & \mbox{ } \hspace{1.5cm}  \times
\delta(\omega+[E^{N_e-1}_m-E^{N_e-1}_0]), 
\label{spinspec}\\  
N(q,\omega)&=&\frac{1}{N}\sum_m|\langle\psi^{N_e-1}_m|N(q)|
\psi^{N_e-1}_0\rangle |^2 \nonumber \\
 &  & \mbox{ } \hspace{1.5cm}  \times
\delta(\omega+[E^{N_e-1}_m-E^{N_e-1}_0]).
\label{chargespec}
\end{eqnarray}
The spin and charge operators, $S(q)$ and $N(q)$ are
\begin{eqnarray}
S(q)&=&\sum_{i}\exp({\rm i}qr_i)(n_{i\uparrow}-n_{i\downarrow}-S_0)
\label{spin} \\ 
N(q)&=&\sum_{i}\exp({\rm i}qr_i)(n_{i\uparrow}+n_{i\downarrow}-N_0)
\label{charge}
\end{eqnarray}
where the summations are over sites, $S_0$ and $N_0$ are the spin 
and charge 
backgrounds ($S_0=(N_\uparrow-N_\downarrow)/N$ and 
$N_0=(N_\uparrow+N_\downarrow)/N$) and 
$n_{i\sigma}=c^\dagger_{i\sigma}c_{i\sigma}$.

The momentum transfer, $q$, is defined via $q=k_F-k$.  
The Fermi momentum, $k_F$ 
is defined as the momentum state with 
the highest single electron energy which is occupied at half-filling
in the non-interacting ($U=0$) case and which 
has positive momentum.  The restriction to $k_F>0$ is important 
in order to 
maintain the association of excitations around the positive 
Fermi point with 
small $q$.  Consequently, for some choices of $\delta$, 
$|\psi^{N_e-1}_0\rangle$ 
in (\ref{spinspec}) and (\ref{chargespec}) is not the true ground state, 
but is the lowest energy level in the subspace of states 
which share the same 
momentum as $c_{k_F\downarrow}|\psi^{N_e}_0\rangle$.  Since both $k$ 
and $k_F$ depend upon $\delta$ according to eq. (\ref{kset}), 
the effect of 
$\delta$ cancels in $q$ as expected ($q$ is a momentum transfer 
and the spacing between the allowed wavenumbers $k$ 
is fixed).

The dynamic spin and charge correlation functions probe the 
properties of the 
excitations in the $N_e-1$ electron system.  A level with a 
strong response in 
$S(q,\omega)$ is taken to indicate a spinon excitation. 
Features in $N(q,\omega)$ are then associated with holon 
excitations.  
Some levels could be due to multiple excitations in one or 
both sets of particles.  
By comparing the spectral function to the corresponding 
spectra of $S(q,\omega)$ 
and $N(q,\omega)$ we have been able to classify many spectral features by 
identifying which excitations are spinon-like and which are holon-like.
This procedure is repeated over the range of $\delta$-values 
$(0\le\delta\le 
2\pi)$. This allows us to plot spectra as a function of a 
quasi-continuous momentum variable, although  strictly
such spectra are taken from an 
ensemble of systems with differing boundary conditions.

Because the HM is integrable in 1D, we can independently
verify our assignments of the features we find in the 
spectral function by reference to solutions to the BA
equations. For an $N$-site Hubbard chain with
$N_e$ particles and $N_{\downarrow}$ down spins and 
a twisted boundary condition these are  ($t=1$) 
\begin{equation}\label{BAcharge}
k_j N = 2\pi I_j + \delta -  \sum_{\alpha=1}^{N_e} 
2 \tan^{-1} \left[ \frac{4(\sin k_i - \lambda_\alpha)}{U} \right]
\end{equation}
and
\begin{eqnarray}
-  \sum_{i=1}^{N_e} 
2 \tan^{-1} \left[ \frac{4(\sin k_i - \lambda_\alpha)}{U} \right]
& = & 2 \pi J_\alpha +
\label{BAspin} \\
& &   \sum_{\alpha = 1}^{N_\downarrow} 
\tan^{-1} \left[ \frac{2(\lambda_\alpha - \lambda_\beta}{U} \right].
\nonumber 
\end{eqnarray}
The quantum numbers ${I_J,J_\alpha}$ characterize the charge
and spin degrees of freedom. 
They are either integer or 
half-odd integer according to the conditions
\begin{eqnarray}
I_j&=&\frac{N_\downarrow}{2}({\rm mod} 1)
\label{quantumI}\\
J_\alpha&=&\frac{N_e-N_\downarrow+1}{2}({\rm mod} 1).
\label{quantumJ}
\end{eqnarray}
The total energy $E$ and total
momentum $P$ of the system of particles is 
\begin{eqnarray}
\label{EBA}
E & = & - 2 \sum_{j=1}^{N_e} \cos k_j  \\
\label{KBA}
P & = & \sum_{j=1}^{N_e} k_j = \frac{2\pi}{N} \left[ 
\sum_{j=1}^{N_e} I_j + \sum_{\alpha=1}^{N_\downarrow} J_\alpha \right]
- \delta \frac{N_e}{N} .
\end{eqnarray}
The BA equations (\ref{BAcharge}) and (\ref{BAspin})
are easy to solve numerically assuming real
rapidities ${k_j,\lambda_\alpha}$ by simple root searching.

\medskip
\section*{III.  Results}
\begin{figure}[h]
  \vspace{0.5cm}
\centerline{
  \epsfxsize=6cm
  \epsfysize=6cm
  \epsffile{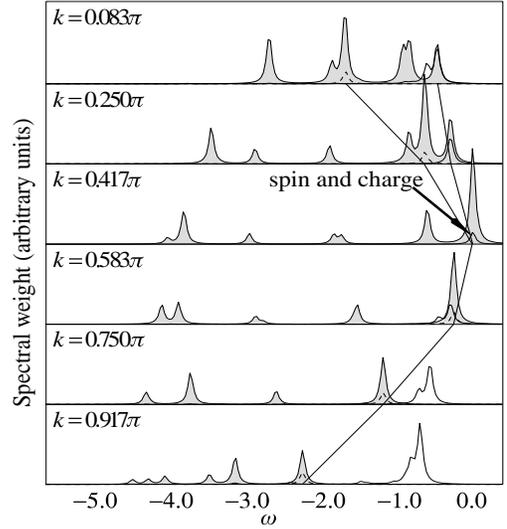}}
  \vspace{1cm}
\caption{\label{raw} The raw spectral functions computed
using Lanczos for a ring of 12 sites at half-filling with 
$U=10$ and $\delta=\pi$.  The spectral
functions (shaded) are shown together 
with the diagnostic spin (continuous line) 
and charge (broken line) correlation functions for the 
11 electron system.  The 
peaks have been Lorentzian broadened to $\Delta\omega=0.04|t|$ and the 
correlation functions have been scaled by a factor 5.  
Some peaks have been joined between different $k$ values with 
thin lines to illustrate the dispersive nature of the principle 
excitations.}
\end{figure}

Figure \ref{raw} is the spectral function for the half-filled 12-site 
Hubbard band
with $U = 10$ and antiperiodic ($\delta = \pi$) boundary conditions
computed using direct diagonalization (Lanczos).  
The curves are cross sections through the spectrum at those 
momenta consistent with this choice of $\delta$.   
The peaks have been Lorentzian broadened as a guide to the eye.  In order 
to label the peaks (spin or charge-like), spectra of the spin and charge 
correlation functions have been superimposed
and the dispersive tendencies of the main spin and charge-like peaks 
have been 
drawn in thin lines.  
The results are very similar to those of Kim et al. \cite{Kim2} 

\mbox{ }

\begin{figure}[htbp]
\vspace{1cm}
\centerline{
  \epsfxsize=6.0cm
   \epsfysize=12.0cm
  \epsffile{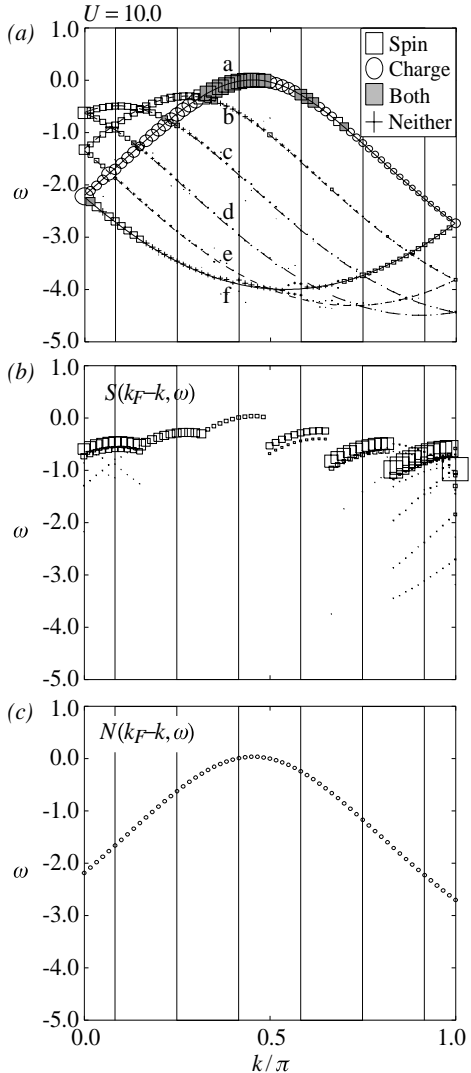}}

\vspace{1cm} 

\caption{\label{classA} (a) The spectral function for 
the 12-site half-filled system with $U=10$ computed by varying the 
boundary condition parameter $\delta$. Peaks in the
spectral function are classified into
`spin-like', `charge-like', `both' and `neither' using the 
spin and charge correlation functions for the 11 particle
system shown in (b) and (c).  
 The areas of the symbols is proportional to the spectral weight in (a) 
 and to the amplitude of the correlation functions in (b) and (c). 
 The continuous curves in 2(a) are those obtained from solving the 
 BA equations and the 
 labeling a-f denotes the quantum numbers $J_\alpha$ shown in
 Table \ref{Jalphas}. } 
\end{figure}

To illustrate the essential features of the 
spectra obtained by exact numerical diagonalization
as $\delta$ is varied,
we  show the spectra 
in the $k$-$\omega$ plane.  Each peak in the spectral function
$A(k,\omega)$ is represented by a symbol centered 
about its momentum ($k$) and 
energy ($\omega$).  Figure \ref{classA}a is 
for the same system as in Figure 
\ref{raw} but for a range of $\delta$. 
The size of the symbols is used to denote the spectral weight while
(as indicated in the inset to  Figure \ref{classA}a)
the shape
 is used to distinguish between spin and charge excitations.
For some peaks denoted by shaded squares, 
there is a contribution from both types of correlation (as indicated 
for example in Figure 
\ref{raw} at $k=0.417\pi$).  The remaining peaks 
which cannot be identified
through the correlation functions are denoted by crosses.  
We also show the spin and charge correlation functions for 
this system (Figures \ref{classA}b and \ref{classA}c respectively) 
which were used to classify the features in 
the spectral function into `charge-like' and `spin-like'. 
(To make the comparison with
$A(k,\omega)$, straightforward $S(q=k_F-k,\omega)$ and $N(q=k_F-k,\omega)$ 
are shown as  
functions of $k$.) 

The vertical lines on Figure \ref{classA} denote those 
momenta corresponding to the choice of $\delta$ in 
(\ref{kset}) which give rise to a closed-shell configuration
for the electrons in the ground state. For the 12 particle system
this corresponds to $\delta = \pi$.  The separation between
two adjacent vertical lines corresponds to a
change of $2\pi$ in $\delta$.

\newpage
\mbox{ }

In Figure 2, the solutions of the BA equations
are superimposed on the results obtained by direct diagonalization. 
Each curve, labeled by 
the letters a-f, corresponds to a different choice of
quantum numbers ${J_\alpha}$ as shown in Table \ref{Jalphas}.
The ${I_J}$ distribution in each case is that of a single
holon, the position of which traces out the complete
holon band as the phase shift is varied. The agreement
between the labeling of the results of the direct diagonalization
and the results of the BA calculation show that the numerical
procedure based on comparison with spin and charge correlation
is a very reliable way of identifying the nature of the 
excitations involved.

In Figure \ref{classA}a there are obvious 
features related to the twisted PBC.
At low binding energies 
and low momenta (upper left part of Figure \ref{classA}a),
there is a series of `humps' in 
the spin-like band, which have an approximate symmetry
about the closed shell lines.   
The main charge-like band, which exists over the full Brillouin 
zone (band `a' in Figure \ref{classA}a), 
has the same shape as one would expect 
to recover in the 
thermodynamic limit.
As we argue in section IV, 
these effects of the  
twisted PBC should be expected and are typical for systems 
with spin-charge separation.

The extension of the charge band to negative momenta (not 
shown) and then reflected about $k=0$ gives the feature which 
disperses with 
positive curvature in the high energy region of Figure \ref{classA}a
(band `f').
Something similar to this curve was seen by Penc et al. \cite{Penc} where
it was interpreted as a shadow of the main holon 
band shifted in momentum by $2k_F$ (we will refer to this band 
as the `shadow band').  Penc et al suggest that this shift is due to 
scattering by spin fluctuations at $2k_F$ (the spin correlation function 
has a 
singularity at $2k_F$ for strongly correlated electrons \cite{Ogata}).  
Most of 
the shadow band is of undefined nature in our results (although some 
peaks do have a very weak spin correlation).  The explanation 
of this feature in 
the context of the Spectral Building Principle
(section IV) is compatible with that of Penc et al.

There are some features in the photoemission spectrum which 
our comparison with the two-particle correlation
function does not clearly identify as being related 
to either spin or 
charge correlations.  These tend to have low spectral 
weight and/or 
to be at energies far from the Fermi energy.  
There are conversely also
features in the correlation functions which are not relevant to 
photoemission.  The spin correlation function 
for example exhibits very strong 
peaks at $k>k_F$ (Figure \ref{classA}b).  These have no photoemission 
association except where they cross the charge band.  
We attribute these humps in the correlation function for $k>k_F$ 
to multiple holon and spinon excitations (section IV).

\begin{table}[htb]  \label{Jalphas}

\begin{tabular}{ccccccc} 
 & \multicolumn{6}{c}{$J_\alpha$}            \\  \hline
a & -2.5 & -1.5 & -0.5 & 0.5 & 1.5 & X \\
b & -2.5 & -1.5 & -0.5 & 0.5 & X & 2.5 \\
c & -2.5 & -1.5 & -0.5 & X & 1.5 & 2.5 \\
d & -2.5 & -1.5 & X  & 0.5 & 1.5 & 2.5 \\
e & -2.5 & X  & -0.5 & 0.5 & 1.5 & 2.5 \\
f &  X & -1.5 & -0.5 & 0.5 & 1.5 & 2.5 \\

\end{tabular} 
\caption{The spin quantum numbers  for the various
solutions (labeled a-f in
Figures \ref{classA} and \ref{classB}) to the
Bethe Ansatz equations (\ref{BAspin}).
 The hole in the spinon band or `antispinon' is denoted
by X.}
\end{table}

The new spinon related structure in the spectral function is a 
finite-size effect.
There are $N/2-1$ spinon cusps present for $-\pi\le 
k<\pi$. For this system and all systems with $N/2$ even,
there is a cusp at $k=0$, 
while for odd $N/2$ the cusp is 
displaced from $k=0$ by $\pi/N$. 
 The range just below $k_F$
in which the spectral function has mixed spin and
charge character is of width $\pi/N$ and becomes
a single point in the thermodynamic limit. 

\begin{figure}[htbp]
\vspace{1.5cm}
\centerline{
 \epsfysize=10cm
 \epsfxsize=6.25cm
  \epsffile{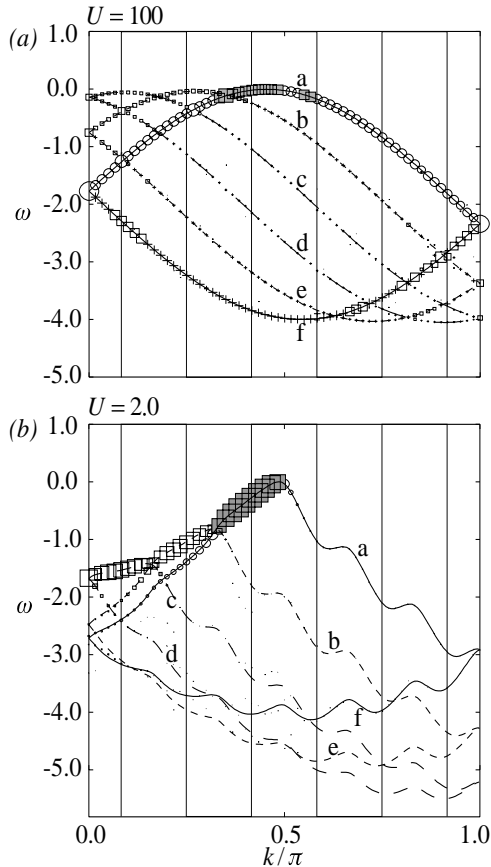}}
\vspace{0.5cm}
\caption{\label{classB} As Figure \ref{classA}a but for $U=100$
and $U=2$.}
\end{figure}

Results for $U=100$  
are shown in Figure \ref{classB}a.
The basic spectral structure of 
Figures \ref{classA}a and \ref{classB}a is similar.  
There 
is more spectral weight in the high energy regions than for $U=10$ with
the shadow band having almost the same weight as
 the main holon band.  The spinon band width 
 parameter $J \sim 4t^2/U$ tends to zero in this limit
 and, in the 
infinite system, the spinon excitations for
$U\to\infty$ do  not 
disperse \cite{Sorella}. Unlike 
the spinon bandwidth, the humps in the spinon part of the spectra 
do not scale with $1/U$.

In the weakly correlated case of $U=2$ shown in Figure
\ref{classB}b, the 
main holon band for $k>k_F$ and the shadow band have a decreasing 
share of the 
spectral weight. (As the 
interaction is reduced, the spectrum is tending towards a 
conventional quasiparticle band with an electron Fermi surface.)  
There is still a two peak structure for 
$k<k_F$, although the identification of 
spin and 
charge-like character of the peaks in Figure \ref{classB}b is only an 
indication of which excitation-type dominates, as
there is notable mixing of 
spin and charge correlations. This is seen in both the
charge and spin correlation functions which show a response 
at the corresponding frequencies.   
The humps in the weak holon band for $k>k_F$ develop into the
one-particle two-hole excitations of the non-interacting
case (where of course they have no spectral weight).

\medskip
\section*{IV. The Spectral Building Principle (SBP)}
The photoemission spectrum 
for a strongly correlated HM in the thermodynamic
limit can be 
explained using the semi-phenomenological 
SBP \cite{Eder1,Kim1}. 
The SBP determines where spectral weight is 
expected and classifies the features as spin or charge-like. 
The SBP does not predict the magnitude of the spectral weight 
itself and this information must be found by other means like exact 
diagonalization or approximate methods.

The SBP assumes that
the ground state of the half-filled system is made up
of a half-filled spinon band containing $N_\downarrow$ spinons 
and an empty 
holon band, with energies $E_s(k_s)$ and $E_h(k_h)$ 
and momenta $k_s$ and $k_h$ for the spinon and holon respectively.  
The SBP treats the spinons and holons as independent excitations 
obeying the Pauli exclusion principle.  
At half-filling the dispersion relations 
are given by \cite{Voit},
\begin{eqnarray}
E_s(k_s)&=&\frac{\pi}{2}J\cos(k_s)
\label{spinon}\\
E_h(k_h)&=&2|t|\cos(k_h).
\label{holon}
\end{eqnarray}

The peaks on the spectral function
are assumed to 
involve a single anti-spinon excitation 
(the annihilation of a spinon from 
the spinon sea) and a single holon excitation 
(the creation of a holon in the 
empty holon dispersion curve). Taking the spinons and holons to be 
independent excitations,
the electron Green's function would then be expected
to be well-described by a convolution of 
the spinon and holon Green's functions.
Conservation of energy and 
momentum would imply
\begin{eqnarray}
\omega&=&E_s-E_h
\label{energy}\\
k&=&k_s-k_h.
\label{momentum}
\end{eqnarray}
Here $\omega$ and $k$, as in Figures \ref{classA}a and \ref{classB}, 
are the energy and momentum lost by the system when an
electron is photoemitted.

According to the SBP, a peak on the spinon band in the spectral function
in the strongly correlated limit is interpreted as a specific 
anti-spinon excitation in combination with the (fixed) lowest energy holon 
excitation at $k_h^0$.  In the thermodynamic limit, $k_h^0=\pi$.  
By varying the 
choice of anti-spinon excitation and using 
(\ref{spinon})-(\ref{momentum}), 
this gives the entire spinon band, 
$\omega_s(k)=(\pi J/2)\cos(k+\pi)+2|t|$  but with 
a cut-off at the spinon Fermi surface at $k=\pm\pi/2$.  In the 
photoemission spectrum, the occupied regions 
of the spinon dispersion have been 
shifted in energy and momentum by the holon excitation at $k_h^0$.  
Likewise, 
the SBP assumes that 
holon bands (main and shadow) are made up of all possible holon 
excitations together with a fixed anti-spinon 
excitation of lowest energy at $k_s^0=\pm\pi/2$, 
corresponding to spinon annihilation at the spinon Fermi points.  
The holon bands in the spectral function have the form, 
$\omega_h(k)=-2|t|\cos(k\mp\pi/2)$.  Since the holon bands measure the 
unoccupied regions of the holon dispersion, the holon band observed
in the photoemission spectrum extends throughout the Brillouin zone.

The SBP as outlined in
\cite{Eder1,Kim1} describes the spectra in the thermodynamic limit.  
In order 
to proceed to finite systems 
and, in particular, to explain the effect of the 
twisted PBC we need to determine how the 
spinons and holons should respond to $\delta$.  
We assume that the holon and spinon
momenta are given by
\begin{eqnarray} 
\label{SBPkh}
k_{hj} & = & \frac{2\pi I_j + \delta}{N}  \\ 
\label{SBPks}
k_{s\alpha} & = & \frac{2\pi J_\alpha}{N} 
\end{eqnarray}
with the $I_j$ and $J_\alpha$ integer or half-integer
as in (\ref{quantumI}) and (\ref{quantumJ}). These quantities
were introduced previously in \cite{Carmelo91} to help
elucidate the connection
between the exact eigenstates of the Hubbard model and the
exact solution of the Tomonaga-Luttinger liquid. We
have generalized them by including the phase shift term in
(\ref{SBPkh}).

The allowed values for the holon quantum numbers ${k_h,k_s}$ or
(equivalently) ${I_j,J_\alpha}$ are functions of the
number of particles $N_e$ and the spin polarization 
$N_\downarrow$. The photoemission process measures the energy
of the $N_e-1$ particle system as a function of the 
missing energy and momenta, so it is natural to expect
the quantum numbers of the $N_e-1$ particle system to characterize
this missing energy and momentum. We therefore assume that
the parity conditions on $I_j$ and $J_\alpha$ are those
for the the $N_e-1$ particle system.

When we compute the spectral function using Lanczos and the energy
dispersion by directly solving the BA equations, we measure
frequency with respect to the ground state of the system
for the same boundary condition. This energy itself depends
on $\delta$ and this effect is not accounted for by the
SBP when we assume the
relation (\ref{energy}). For $U=2$, the variation
in the ground state energy of the half-filled case is
less than 10\% of the spinon bandwidth and is smaller
for larger $U$. For the
spectral function taken for systems which are not
half-filled and with no gap the picture is more complicated
and will be addressed in a later publication.

\begin{figure}[htbp]
\vspace{0.5cm}
\centerline{
  \epsfxsize=7.5cm  
  \epsfysize=4.97cm
  \epsffile{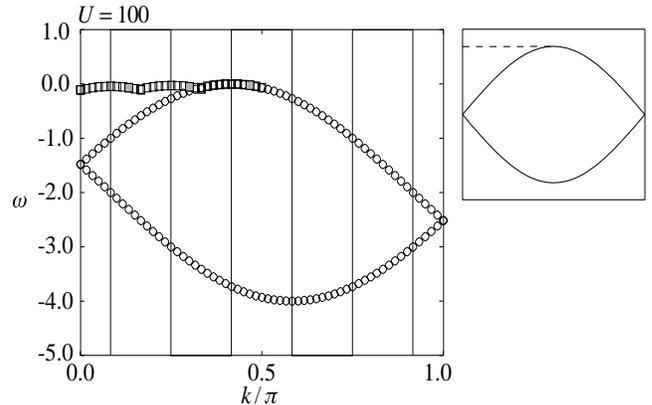}}
  \vspace{0.5cm}
\caption{\label{SBPlargeU} The main picture (left) shows the 
photoemission 
spectrum for $N=12$, $N_e=12$, $N_\downarrow=6$ and $U=100$ 
computed on the 
basis of the SBP (squares=spinon, circles=holon) and should 
be compared with the 
result of Figure \ref{classB}a.  The shaded points are for reference 
in the text 
to follow.  The small plot shows the corresponding result in 
the thermodynamic 
limit (continuous line=charge, broken line=spin).}
\end{figure}

Figure \ref{SBPlargeU} shows the positions of the spinon and holon peaks
computed using the SBP for the case $U=100$
and should be compared with the numerical 
results of Figure \ref{classB}a. 
The spinon regions are denoted by squares and the 
holons with circles.  The agreement is good bearing in mind that 
only the spectral 
features which are associated with $k_h^0$ and $k_s^0$ are reproduced in 
Figure \ref{SBPlargeU}).  The main holon band, the shadow 
band and the spinon 
cusps are reproduced.  The number of cusps and their positions agree 
well with 
the numerical result and this picture also gives an offset of the 
top-most 
energy of the holon band to a slightly lower momentum than $k_F$ 
which is 
evident in the numerical result too.  
Shown for comparison is the result for the 
thermodynamic limit where there are no cusps and the top of the holon 
band is at 
$k_F$ exactly.  

\begin{figure}[htbp]
\vspace{1cm}
\centerline{
  \epsfxsize=7.5cm 
  \epsfysize=4.5cm
  \epsffile{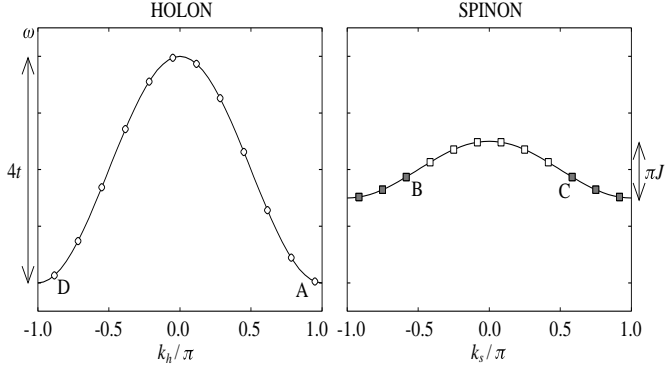}}
\vspace{0.5cm}
\caption{\label{shDisp} The holon and spinon dispersions for the finite system 
$N=12$, $N_e=12$, $N_\downarrow=6$ and $\delta=1.6\pi$.  Open (filled) symbols 
denote unoccupied (occupied) holon and spinon states.  The allowed holon states 
change with $\delta$ but those of the spinon are static.}
\end{figure}

The shape of the spectra Figure \ref{SBPlargeU} can be understood
by considering the 
holon and spinon dispersions for a given $\delta$ shown in 
Figure \ref{shDisp}. For the spinon 
band in Figure \ref{SBPlargeU} the set of occupied states of the spinon 
dispersion are all shifted together by the lowest energy holon 
excitation---labeled `A' in Figure \ref{shDisp} (note that the 
spinon bandwidth is exaggerated in Figure \ref{shDisp}).  According to  
(\ref{spinon}) and (\ref{holon}) this gives one set of equally spaced points 
on the spinon band in Figure \ref{SBPlargeU} (the shaded points).  Varying 
$\delta$ changes both the energy and momentum of the lowest energy holon
while the spinon energy is (by assumption) independent of $\delta$. As
a result, each peak in the spinon band of the 
 spectral function is shifted to a 
slightly different position. 
The humps observed in the spinon band of the spectral function then
simply reflect the variation with $\delta$ of the holon state
labeled by `A'.
At certain values of $\delta$ the lowest energy holon excitation changes 
its momentum discontinuously as the state previously labeled by 
`A' moves up the holon band.  Whenever this occurs it gives 
rise to the cusps in the spinon 
band in the spectral function.

The form of holon bands in the spectral function can also be understood
using the SBP.  The shift of the holon dispersion by the spinon
excitation `B' gives rise to the main holon band (the upper holon band
in Figure \ref{SBPlargeU}).  Since (by assumption) the allowed momenta
in the spinon dispersion are independent of $\delta$, the holon
dispersion is shifted by a constant amount as $\delta$ is varied.  This
is the reason why the holon bands in the numerical spectra are smooth
and free of cusps in the strongly correlated limit.  The alternative
spinon excitation, `C', has the same energy as `B' but shifts the holon
dispersion by a different momentum.  This leads to the so-called shadow
band in the spectral function (lower holon band in Figure
\ref{SBPlargeU}).  The momentum difference between `B' and `C' is $2k_F$
(true for all fillings) which explains why the shadow band is displaced
by $2k_F$ from the main holon band.

The coupling between spin and charge which exists
for moderate $U$ means that
assuming completely independent spinon
and holon bands as in (\ref{spinon}) and (\ref{holon}) 
is no longer strictly valid. Even so it is still possible to 
explain the spectra
using the SBP, although the non-zero spinon bandwidth $J$
introduces a minor complication as we explain below.

\begin{figure}[htbp]
\vspace{0.5cm}
\centerline{
  \epsfxsize=7cm
  \epsfysize=3.5cm
  \epsffile{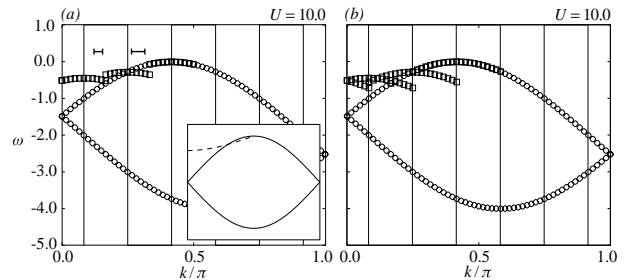}}
  \vspace{0.5cm}
\caption{\label{SBPmoderateU} Spectral function computed using the spectral 
building principle for $N=12$, $N_e=12$, $N_\downarrow=6$ and $U=10$.  In (a) 
the spinon (holon) band is shifted by the very lowest 
energy holon (anti-spinon) 
excitations only.  The inset is for the thermodynamic limit (continuous line = 
charge, broken line = spin).  In (b) we allow the spinon dispersion to be 
shifted by the next lowest energy holon excitations also.  This should be 
compared with Figure \ref{classA}a.}
\end{figure}

If we implement the SBP exactly as described above for the large
$U$ case, the low energy peaks predicted by the SBP do not
agree completely with what we find using direct diagonalization
and the BA solution.
Figure \ref{SBPmoderateU}a shows the 
photoemission spectrum computed on the basis of the SBP 
for $U=10$ ($J=0.4$) and $N=12$ at 
half-filling (shown in the inset is the $N\to\infty$ result for comparison).  
Unlike the numerical result of Figure \ref{classA}a there appear to be
discontinuities the spinon band while the holon bands look
very similar to what we found numerically  (see Figure \ref{classA}a).
These discontinuities come about because we have assumed that,
as in the large $U$ case,
the spinon band of the photoemission spectrum 
corresponds to the combination of 
a spinon with the {\it lowest} energy holon excitations as 
before. Whereas this is correct for most
momenta (with $k\le\pi/2$) in Figure \ref{SBPmoderateU}a,
it is not true for the range of momenta 
highlighted by the horizontal `bars' in the top-left of Figure 
\ref{SBPmoderateU}a.
We find that for these 
electron momenta, lower energy electronic states can be formed 
by combining a spinon excitation 
 with the slightly {\it higher} energy holon excitations at point `D' 
in Figure \ref{shDisp}.
Figure \ref{SBPmoderateU}b shows the photoemission spectrum 
computed using the SBP adjusted to allow also for low energy
spinon and holon combinations ({\it ie\/} using states
at `D' as well as `A' in Figure \ref{shDisp}).
Although it might seem
peculiar that an electronic state
formed from a higher energy holon state (`D') can be
responsible for lower energy electronic states, it can be explained
as follows. The holon at point `D' in Figure \ref{shDisp} ($k_h^D,E_h^D$) 
has a higher 
momentum than `A' ($k_h^A,E_h^A$) (after adding $2\pi$ to $k_h^D$).  
To form a hole with given momentum,
the spinon state combined with the holon state at
`D' will be at a different point on the spinon band. 
The cost in energy of using the more energetic holon state `D' can then be
compensated by a lower spinon energy.  This adjustment is not 
necessary for the $U\to\infty$ case since there the spinon band is flat.

\mbox{ }

Although $U=2$ does not represent a weakly correlated system, it is too 
small for the SBP to work well.  We find, in fact, that 
the SBP ceases to work well 
below $U\approx 4$, where the holon band starts to
develop humps as a function of $\delta$. We have tracked these humps
in the exact BA solution
as a function of $U$ down to small $U$ ($U<0.1$). They develop
into the two-hole and one-particle excitations of 
the non-interacting system,
where of course they have vanishing spectral weight.

In the SBP plots (Figures \ref{SBPlargeU} and \ref{SBPmoderateU}), we have 
shown only those peaks which are firmly classified using the
results of the spin and charge correlation functions. These
turn out to those 
associated with the lowest (or in some cases low-lying) spinon or 
holon 
excitations.  The numerical spectra contain many other peaks 
in addition to the 
spinon and holon bands.  By taking every possible combination of 
{\it single\/} 
holon and anti-spinon excitations - regardless of energy - the SBP can
reproduce essentially all the 
peaks in the spectral function.  It is clear then that the 
spectral function are dominated by the creation of 
{\it single\/} holon plus anti-spinon 
excitations.  

Other classes of excitation exist which can be 
accounted for using the SBP but which are invisible in the 
spectral function. 
 For example for the $U=10$ system of Figure \ref{classA} it
is clear that that there are contributions 
to the spin correlation function for $k>k_F$  for 
momenta $k>\pi/2$ (there is no significant charge correlation in addition to 
that already discussed).  
We attribute
these `extra' humps to multiple excitations, 
namely the holon and 
anti-spinon excitations already described accompanied by a 
spinon/anti-spinon pair (a spin wave). The effect of the twisted
PBC makes these easy to identify,
as the spinon humps at $k>k_F$ appear as
images of one or more of the original  humps in the spinon band
but shifted by the extra momentum 
and energy of the spinon/anti-spinon pair. 
The spin waves can provide potentially any 
momentum in the set $\{2n\pi/N\}$ and an energy which is finite but small (due 
to the narrow width of the spinon band).

\medskip
\section*{V. Summary}
We have studied the photoemission spectrum of 1D Hubbard
model at half-filling as a function of the periodic boundary
condition (PBC) in finite size chains. We have shown
that the finite-size effects can be well explained
using 
the spectral building principle (SBP, \cite{Kim1,Eder1})
adapted to the case of finite systems and the twisted PBC,
in terms of independent populations of spinons and holons.
We have found that the photoemission spectral functions are
explained for $U>\sim 4$ by assuming that only the holons are 
directly dependent
on the twisted PBC and that one holon-one antispinon pairs 
account for essentially all the spectral weight. Other excitations
involving additional spin-wave processes
are observed in the spin correlation function.

The key feature found in the strongly
correlated systems ($U>\sim 4$), 
in which the spinons and holons may
be thought of as independent excitations,
are the `humps' in the spinon bands of the photoemission
spectra. We have shown that 
the curvature of these humps directly reflects
the curvature of the holon dispersion at the bottom of the 
holon band and so could be used 
to extract directly the holon effective mass in cases where
there is no direct independent way to compute this quantity.

We have restricted our study to the 1D nearest neighbor Hubbard
model where we have been able to
show that identifying the spin and charge excitation bands
using the corresponding correlation functions
agrees exactly with the results from Bethe Ansatz exact calculations.
These results suggest that this method can be extended to other 
1D systems and higher dimensional systems
for which there is no exact BA solution for the model \cite{Bannistertogo}.

We wish to thank the EPSRC for providing the funding for this work.

\medskip
\bibliographystyle{plain}

\widetext

\end{document}